\documentclass[twocolumn,prb,aps,showpacs,superscriptaddress,showkeys,reprint]{revtex4-1}

\usepackage{amssymb, amsmath}
\usepackage[english]{babel}
\usepackage{bm}
\usepackage{graphicx}
\usepackage[utf8]{inputenc}
\usepackage{color}

\newcommand{\hlchanges}[1]{{#1}}

\begin{document}
\author{Marten Richter}
\email[]{marten.richter@tu-berlin.de}

\affiliation{Institut für Theoretische Physik, Nichtlineare Optik und
Quantenelektronik, Technische Universität Berlin, Hardenbergstr. 36, EW 7-1, 10623
Berlin, Germany}

\title{Nanoplatelets as material system between strong confinement and weak confinement}


\begin{abstract}
 Recently, the fabrication of CdSe nanoplatelets became an important research topic.
 Nanoplatelets are often described as having a similar electronic structure as 2D dimensional quantum wells and are promoted
 as colloidal quantum wells with monolayer precision width. In this paper, we show, that nanoplatelets are not ideal quantum wells, but cover depending on the size: the strong confinement regime, an intermediate regime and a Coulomb dominated regime.
 Thus, nanoplatelets are an ideal platform to study the physics in these regimes.
 Therefore, the exciton states of the nanoplatelets are numerically calculated by solving the full four dimensional Schrödinger equation.  We compare the results with approximate solutions from semiconductor quantum well and quantum dot theory.
 The paper can also act as review of these concepts for the colloidal nanoparticle community. 
 \end{abstract}


\date{\today}
\maketitle


In quantum dots the wave function of the electron and hole, that form the optically created exciton are confined in all three dimensions resulting in a quasi zero dimensional system with discrete states \cite{Brus:IEEEJQuantElect:86,Efros:PhysRevB:96,Bimberg::99,Rajh:JPhysChem:93,Eychmuller:ChemPhysLett:93}.
The chemical synthesis of colloidal quantum dots is a very active research field \cite{Anikeeva:NanoLett:07,Mahler:Naturematerials:08,Sargent:Naturephotonics:12,Shirasaki:NaturePhotonics:13}, 
since the controlled growths of these materials lead to many real life applications  like dyes  for e.g. television \cite{Shirasaki:NaturePhotonics:13}  or as markers in biology \cite{Medintz:Naturematerials:05}.
Colloidal quantum dots may have a higher potential for applications than epitaxial grown quantum dots, which are more difficult to grow \cite{Shirasaki:NaturePhotonics:13}.

On the other hand, epitaxial grown structures like quantum wells serve very well in many lighting applications like LEDs \cite{Nakamura:JapanJApplPhys:95,Tan:IEEEPhotonicsJournal:12,Nakamura:RevModPhys:15}. 
In quantum wells the electron and hole are  only in one dimension confined and can move freely two dimensions.
Optical excitation creates in these quasi 2D system bound electron-hole pairs: excitons \cite{Haug::94,Zimmermann::03,Zimmermann:PureApplChem:97}.
Colloidal nanoplatelets (e.g. CdSe) can be grown with monolayer precision \cite{Ithurria:JACS:08,Pelton:NanoLetters:12,Achtstein:NanoLetters:12,Tessier:ACSNano:12,Benchamekh:PhysRevB:14,Pelzer:JChemPhys:15,Naeem:PhysRevB:15,Achtstein:PhysRevLett:16,Cassette:PhysChemChemPhys:17}, therefore nanoplatelets may be
a chemical grown alternative to epitaxial quantum wells.
However sizes of nanoplatelets are not as big as quantum well sizes. Therefore the size of most nanoplatelets is not large enough for their exciton states to have the same properties like a quantum well. 
On the other hand most nanoplatelets are too large for quantum dot like properties and photoluminescence spectra show similar features as quantum wells \cite{Ithurria:JACS:08,Tessier:ACSNano:12,Achtstein:NanoLetters:12,Achtstein:PhysRevLett:16}, since disorder in quantum wells like confinement within the platelet area  usually leads to a similar localization of states.

In the moment, the toy models used to describe the optical relevant exciton states for nanoplates depend often on the scientific background of the authors: Some paper use an exciton wavefunction factorized in relative motion of electron and hole and the center of mass motion of the whole exciton \cite{Naeem:PhysRevB:15}. This is the correct approach for a quantum well with disorder, if the Coulomb interaction between electron and hole dominates compared to the confinement/disorder. 
Other papers from authors with a quantum dot background \cite{Achtstein:PhysRevLett:16} use the ansatz for the strong confinement limit, where the exciton wavefunction is factorized into an electron and an hole part. This is the correct approach, if confinement dominates compared to Coulomb interaction.
Treatments with a chemical background \cite{Pelzer:JChemPhys:15} used a Frenkel exciton ansatz, which does not reflect the properties of the more Wannier type excitons in the nanoplatelets.

A detailed look at the typical model ansatz wavefunctions for platelets, previously known  from quantum wells and dots, is necessary and is provided in this paper (this paper complements very early studies \cite{Bryant:PhysRevB:88} using a variation ansatz focussed on the ground state, with a full solution of the four dimensional Schrödinger equation and an analysis of higher excited states).
The study should also complement the information obtained through recent ab initio studies using periodic arrangements in inplane direction\cite{Pal:NanoLetters:17,Benchamekh:PhysRevB:14}, since we use finite sizes in inplane direction.
We start with the derivation/introduction of a Wannier type model system for obtaining the four dimensional excitons wave functions.
The four dimensional wave functions were  numerically calculated using finite differences for various nanoplatelet sizes as basis for an analysis of the exciton states. The interplay between Coulomb coupling and confinement is the main key to understand the exciton state properties: therefore the approaches for the strong confinement regime and the Coulomb dominated regime are discussed after this. The full solution and the two approximations are then compared and discussed for nanoplatelets of different size covering different regimes.

\section{Model system: the four dimensional Wannier equation}
 \begin{figure}[htb]
  \includegraphics[width=4cm]{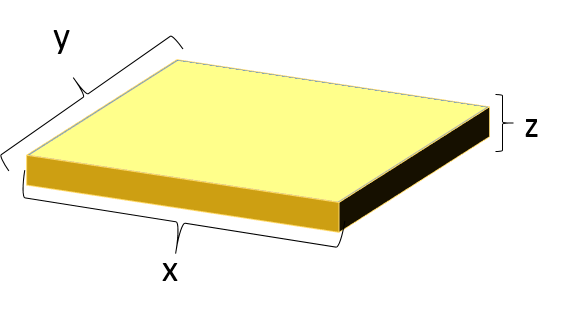}
  \caption{Sketch of a nanoplatelet.}
  \label{scheme}
 \end{figure} 
The main aim of this paper is to understand the interplay of confinement and Coulomb coupling in nanoplatelets for choosing the right model system in an analysis. For a qualitative understanding the model should be as simple as possible, on the expense of quantitative accuracy.
Platelets are box-like nanostructures, which are typically in z-direction only a few monolayers thin, while in x- y- direction their extent is much larger (cf. Fig. \ref{scheme}).
\hlchanges{ For describing the wavefunction of the platelet excitons, a standard multiband envelope ansatz can be used \cite{Haug::94, Zimmermann::03,Ekimov:JOSAB:93}:
$\mathbf{\Psi}(\mathbf{r}_e,\mathbf{r}_h)
=\sum_{\lambda_e\lambda_h}\sum_{ n_e,n_h}  \Psi_{\lambda_e \lambda_h}^{n_e,n_h}(\bm{\rho}_e,\bm{\rho}_h)\zeta_{n_e}(z_e) \zeta_{n_h}(z_h) u_{\lambda_e}(\mathbf{r}_e) u_{\lambda_h}(\mathbf{r}_h)$ with $\mathbf{r}_i=(\bm{\rho}_i,z_i)$. Here, $\lambda_i$ ($n_i$) are the (sub-)band indices for the electron (e) and hole (h) of the exciton.  $u_{\lambda}(\mathbf{r})$ are the Bloch functions around the band edge $\mathbf{k}\approx0$ (cf. Ref. \onlinecite{Haug::94}) describing the wavefunction on scales below unit cell size.
The envelope function $\Psi_{\lambda_e \lambda_h}^{n_e,n_h}(\bm{\rho}_e,\bm{\rho}_h)$ describes the exciton inplane motion and the envelope functions $\zeta_{n_i}(z_i)$ describe the carrier $z$ motion on scales above the unit cell sizes.}
\hlchanges{ Since} CdSe is a wide gap semiconductor, \hlchanges{where } band mixing effects are not dominant, we \hlchanges{can} choose a simple parabolic two band model for the electron and hole carriers instead of a multiband approach like the Kane model \cite{Kane:JPhysChemSolids:57,Norris:PhysRevB:96} to keep the discussion simple. \hlchanges{Furthermore, only  the lowest electron and hole subband are considered, since we are interested in the exciton states around the band edge. }
\hlchanges{This leads to the  exciton wavefunction ansatz (cf. Ref. \onlinecite{Zimmermann:PSSa:97}): $\mathbf{\Psi}(\mathbf{r}_e,\mathbf{r}_h)
=  \Psi(\bm{\rho}_e,\bm{\rho}_h)\zeta_{e}(z_e) \zeta_{h}(z_h) u_{e}(\mathbf{r}_e) u_{h}(\mathbf{r}_h)$.} \hlchanges{ Instead of using envelope functions $\zeta_{i}(z)$ also tight binding approaches can be used to describe the wavefunction z-direction part and the Bloch part (like in Ref. \onlinecite{Benchamekh:PhysRevB:14})}.
\hlchanges{In all these approaches (including the limit of an infinite thin platelet), the starting point for further discussion is the four dimensional stationary Schrödinger equation of the envelope function for inplane motion $\Psi(\bm{\rho}_e,\bm{\rho}_h)$ }:
$( -\frac{\hbar^2}{2 m_e} \Delta_{\bot e} -\frac{\hbar^2}{2 m_h} \Delta_{\bot h}
 +V_{c,e}(\bm{\rho}_e)+V_{c,h}(\bm{\rho}_h)+U_{coul}(\bm{\rho}_e-\bm{\rho}_h))\Psi(\bm{\rho}_e,\bm{\rho}_h)=E \Psi(\bm{\rho}_e,\bm{\rho}_h) $\footnote{\hlchanges{Only constant prefactors  in quantities like the optical dipole element will slightly differ for the different approaches.}}.
The first two terms describe the motion of the electron (hole) with the effective masses $m_e$ and $m_h$ in the conduction and valence bands \cite{Haug::94,Zimmermann::03,Zimmermann:PureApplChem:97}.
The independent free motion of the electron and holes is an important property, that distinguishes Wannier excitons in inorganic semiconductors from Frenkel excitons in organic semiconductors.  Please note, a model with excitons at different sites like in the Frenkel case with coupling between the excitons \cite{Pelzer:JChemPhys:15} is not sufficient for describing this inorganic semiconductor nanostructure, since electron and holes can move independently to some extent determined by Coulomb interaction.
Instead in the case of a site model independent tunneling Hamiltonians for electron and holes like in Ref. \onlinecite{Meier::06} are required to include the independent motion of electron and holes in the two bands.
$V_{c,e}(\bm{\rho}_e)$ and $V_{c,h}(\bm{\rho}_h)$ are the confinement potentials of the electron and hole.
In order to keep the model system simple, an infinite deep confinement potential is used with
$V_{c,e/h}(\bm{\rho}_{e/h})=0$ for $\bm{\rho}_{e/h}$ inside the platelet and 
$V_{c,e/h}(\bm{\rho}_{e/h})=\infty$ for $\bm{\rho}_{e/h}$ outside of the platelet.                            
The actual confinement potential may be smooth and have finite depth in reality. However, e.g. for quantum wells,  a finite depth potential well can be approximated with an infinite potential well with a smaller effective size. Therefore, the platelets from the model system with the infinite potential well are slightly smaller as the platelets in reality.
$U_{coul}(\bm{\rho}_e-\bm{\rho}_h)$ is the Coulomb potential between the electron and holes.
In atomically thin 2D materials, the Coulomb potential is  modified \cite{Keldysh:SovJExpTheoPhysLett:79,Berkelbach:PhysRevB:13} and described by the asymptotic approximation of the potential \cite{Berkelbach:PhysRevB:13} $
U_{coul,K}(\bm{\rho})=1/(4\pi\epsilon_0(2 \epsilon_{r,out})\rho_0) (\mathrm{ln}(\rho/(\rho+\rho_0))+(\gamma-\mathrm{ln}(2))e^{-\rho/\rho_0})$ with $\rho_0=z_0 \epsilon_r/(2\epsilon_{r,out})$,
with the platelet thickness $z_0$ and $\varepsilon_{r}$ of the platelet material and $\varepsilon_{r,out}$ of the solvent that surrounds the platelet and the euler constant $\gamma$.
Platelets are only a few monolayers thin, so in this respect the potential may describe platelets well, \hlchanges{ and may be better than using a vacuum Coulomb potential together with envelope functions in $z$ direction - the standard approach for quantum wells} \cite{Miller:PhysRevB:85,Mayrock:PhysRevB:99,Chow:PhysRevB:99}.
However, their size in the perpendicular direction is not  large enough, that the assumption of an infinite two dimensional material holds. The approximation is better suited for larger platelets, but it is a good first rough approximation.
The potential $U_{coul,K}(\bm{\rho})$ has a logarithmic singularity, for a calculation using finite differences a very high discretization is required. This is not numerically feasible for the full four dimensional problem.
 Therefore the parameter $\alpha_0$ of a model potential \cite{Mayrock:PhysRevB:99} $
 U_{coul}(\bm{\rho})=-1/(4\pi\epsilon_0 \epsilon_r) 1/(\sqrt{{\rho}^2+(\alpha_0 z_0)^2)})$ \hlchanges{is adjusted},
so that the binding energies of the lowest energy eigenstates  for eigenfunctions of the relative wavefunction of an infinitely extended platelet   match approximately the corresponding eigenenergies of the Keldysh potential  $U_{coul,K}(\bm{\rho})$.\footnote{The error for the first excited state binding energy is below 2\% and the error for second and third excited state binding energy is below 20 \%.}
This approach is similar to the procedure used in Ref. \onlinecite{Mayrock:PhysRevB:99}, and allows to the describe the Coulomb induced correlated motion of electron and hole. 
The approximation allows a qualitative discussion, but the quantitative exciton energies of the higher excited states should be discussed with care.  
The numerical solution of the full four dimensional problem is calculated using the effective potential $U_{coul}(\bm{\rho})$. For a consistent discussion all approximate solutions use the same effective Coulomb potential.

\hlchanges{The interband optical transition strength for creating excitons
is determined by the product of overlap integrals $\int\mathrm{d} z \zeta_{e}(z) \zeta_{h}(z)$ of the envelopes $\zeta_{e}(z_e)$ and $\zeta_{h}(z_h)$, the interband dipole (or momentum) transition element $\bm{d}_{cv}$ ( $\bm{p}_{cv}$) and overlap integral between electron hole in inplane direction $O_{n}=\int\mathrm{d}^2 \rho \Psi_n(\bm{\rho},\bm{\rho})$ (see Refs. \onlinecite{Haug::94,Banyai::93,Zimmermann:PSSa:97,Zimmermann::03,Gu:PhysRevB:13}).}
\hlchanges{In order to quantify the oscillator strength for creating the exciton states with different envelope functions, the relevant overlap integral between electron and hole: alone $O_{n}$ is sufficient for the following analysis in this paper, since we focus on the exciton states around the band edge}.
We used the following parameters 
$z_0=0.302 ~\textrm{nm} \times4.5~ \textrm{monolayers}$ \cite{Achtstein:NanoLetters:12}, 
$m_e=0.22$, 
$m_h=0.41$ (Ref. \onlinecite{Benchamekh:PhysRevB:14}),
$\alpha_0=1.1$, $\epsilon_r=9.5$, $\epsilon_{r,out}=5.0$ for the calculations in this paper.

Before the discussion of bound exciton eigenstates of the full four dimensional Schrödinger equation  the strong confinement limit and the weak confinement limit are recapitulated.

\subsection{Strong confinement limit}
Electrons and holes of an exciton state are correlated by Coulomb interaction.
In the strong confinement limit\cite{Efros:SovPhysSemi:1982,Grundmann:PhysRevB:95} the confinement potential restrains the motion of electron and holes on a smaller distance as the typical distance between electron and holes in a bound exciton states in the unconfined case.
Though the shape of the wavefunction  is almost completely determined by the confinement potential and is  not influenced by the Coulomb potential in a first approximation. 
\hlchanges{So that} the four dimensional Schrödinger equations can be approximated in zeroth order as\cite{Efros:SovPhysSemi:1982}:
$( -\frac{\hbar^2}{2 m_e} \Delta_{\bot e} -\frac{\hbar^2}{2 m_h} \Delta_{\bot h}
 +V_{c,e}(\bm{\rho}_e)+V_{c,h}(\bm{\rho}_h))\Psi(\bm{\rho}_e,\bm{\rho}_h)=E \Psi(\bm{\rho}_e,\bm{\rho}_h)$. 
The factorized ansatz  $\Psi(\bm{\rho}_e,\bm{\rho}_h)=\psi_e(\bm{\rho}_e) \psi_h(\bm{\rho}_h)$ yields two equations one for the electron $( -\frac{\hbar^2}{2 m_e} \Delta_{e\bot} 
 +V_{c,e}(\bm{\rho}_e))\psi_e(\bm{\rho}_e)=E_e \psi_e(\bm{\rho}_e)$ 
and one for the hole  $( -\frac{\hbar^2}{2 m_h} \Delta_{h\bot} 
 +V_{c,h}(\bm{\rho}_h))\psi_h(\bm{\rho}_h)=E_h \psi_h(\bm{\rho}_h)$ wavefunction.
The exciton eigenstates are described by an electronic eigenfunction $\psi_{e,n}(\bm{\rho}_e)$ and an hole eigenfunction  $\psi_{h,m}(\bm{\rho}_h)$.
The overlap between electron $n$ and hole wavefunction $m$ \cite{Banyai::93}:
$ O_{eh,nm}=\int \mathrm{d}^2\rho \psi_{e,n}^*(\bm{\rho}) \psi_{h,m}(\bm{\rho})$
\hlchanges{enters the dipole strength as a factor and thus determines} if an exciton composed of electron $n$ and hole $m$ is a dark or a bright state.
The exciton has the energy $E_{nm}^s=\varepsilon_n+\varepsilon_m$ relative to the band gap.
Even if Coulomb coupling does not determine the shape of the exciton wave function in the strong confinement limit, it shifts the exciton energy considerably.
The  Coulomb shift can be obtained in first order perturbation theory by \cite{Grundmann:PhysRevB:95}:
$ V_{Coul,nm}=\int\mathrm{d}^2\rho_e \int\mathrm{d}^2\rho_h  |\psi_{e,n}(\bm{\rho}_e)|^2 U_{coul}(\bm{\rho}_e-\bm{\rho}_h) |\psi_{h,m}(\bm{\rho}_h)|^2$,
giving the Coulomb corrected exciton energy ${E'}_{nm}^s =\varepsilon_n+\varepsilon_m+V_{Coul,nm}$.
Higher order Coulomb correction, e.g. using second order perturbation theory \cite{Grundmann:PhysRevB:95} or Hartree-Fock equations \cite{Achtstein:PhysRevLett:16}, are also used in the literature and may extend the validity range of the strong confinement model towards larger platelets.
We will see in the numerical analysis, that the strong confinement limit describes small nanoplatelets very well.

\subsection{Weak confinement: Coulomb dominated limit}
If the confinement area is large, 
the Coulomb attraction between electron and holes creates bound exciton states, showing a correlated electron and hole motion.
Exciton states with a higher binding energy have a lower average distance between electron and holes.
The first step to attack the weak confinement limit is to ignore the confinement potential and to solve the Schrödinger equation including only  Coulomb interaction beside the kinetic terms:
$( -\frac{\hbar^2}{2 m_e} \Delta_{\bot e} -\frac{\hbar^2}{2 m_h} \Delta_{\bot h}
+U_{coul}(\bm{\rho}_e-\bm{\rho}_h))\Psi(\bm{\rho}_e,\bm{\rho}_h)=E \Psi(\bm{\rho}_e,\bm{\rho}_h)$.
The problem without the confinement potential is translational invariant, a formulation using relative coordinates $\mathbf{r}=\bm{\rho}_e-\bm{\rho}_h$
and center of mass coordinates $\mathbf{R}=m_e \bm{\rho}_e+m_h \bm{\rho}_h$ is beneficial\cite{Haug::94,Zimmermann::03}: $
  ( -\frac{\hbar^2}{2 m_r} \Delta_r -\frac{\hbar^2}{2 M} \Delta_R
+U_{coul}(\mathbf{r}))\Psi(\mathbf{r},\mathbf{R})=E \Psi(\mathbf{r},\mathbf{R})$
with the reduced mass $1/m_r=1/m_h+1/m_e$ and overall mass $M=m_e+m_h$.
Using the factorization Ansatz $\Psi(\mathbf{r},\mathbf{R})=\psi_r(\mathbf{r})\psi_{COM}(\mathbf{R})$,
we obtain the Wannier equation for the relative wavefunction:
$( -\frac{\hbar^2}{2 m_r} \Delta_r 
+U_{coul}(\mathbf{r}))\psi_r(\mathbf{r})=E_r \psi_r(\mathbf{r})$.
The Coulomb interaction between electron and holes in the Wannier equation creates bound states of the relative wavefunction, analog to the hydrogen atom. The mean distance between electron and hole of a particular state compared to the confinement dimensions is an indicator, if the factorization in relative and center of mass wavefunction is a good approximation. For the lowest energy eigenstate, the average distance is not far away from the exciton Bohr radius $a_B$ (factor $1.5$ for the ideal hydrogen 1s state).
  Of course the exciton Bohr radius alone is not a sufficient  indicator for higher energy eigenstate with bigger radii (see later the discussion of states including 2s and 2p and higher contributions).
The confinement energy compared to the binding energy is also a good indicator, whether the strong or weak confinement is the correct limit.
For a selected eigenfunction $\psi_{r,n}(\mathbf{r})$ of the Wannier equation, the full Schrödinger equation can be used for obtaining an equation for obtaining the wavefunction of the center of mass motion (COM)\cite{Haug::94,Zimmermann::03}:
$( -\frac{\hbar^2}{2 m_r} \Delta_r -\frac{\hbar^2}{2 M} \Delta_R
+U_{coul}(\mathbf{r})+V_{c,e}(\mathbf{R}+\frac{m_h}{M}\mathbf{r})+V_{c,h}(\mathbf{R}-\frac{m_e}{M}\mathbf{r}))\psi_{r,n}(\mathbf{r})\psi_{COM,n}(\mathbf{R})=E \psi_{r,n}(\mathbf{r})\psi_{COM,n}(\mathbf{R})$.
Multiplying the equation by $\psi_{r,n}^*(\mathbf{r})$ and integrating over $\mathbf{r}$ yields\cite{Zimmermann::03}:
 $( -\frac{\hbar^2}{2 M} \Delta_R
+\tilde{V}_{c,e,n}(\mathbf{R})+\tilde{V}_{c,h,n}(\mathbf{R}))\psi_{COM,n}(\mathbf{R})=(E-E_{r,n}) \psi_{COM,n}(\mathbf{R})$.
Here we introduced  the effective confinement potentials \cite{Zimmermann:PSSa:97}
$\tilde{V}_{c,e/h}(\mathbf{R})=\int\mathrm{d}^2r V_{c,e/h,n}(\mathbf{R}\pm\frac{m_{h/e}}{M}\mathbf{r}) |\psi_{r,n}(\mathbf{r})|^2$.
 We replace the potential $\tilde{V}_{c,e/h}(\mathbf{R})$ with the confinement potential of the electron and holes as a first approximation.
Another approximate way would be using a confinement potential for the COM wave function reduced in its size by the averaged diameter of the relative wave function, however this does not work for structures smaller or around the diameter of the relative wave function.
The full wave function in center of mass approximation $\Psi_{n,m}(\mathbf{R},\mathbf{r})=\psi_{r,n}(\mathbf{r}) \psi_{COM,n,m}(\mathbf{R})$ is characterized by the quantum numbers $n$ of the relative wave function and by the quantum numbers $m$ of the COM motion.
It is important to note that in general the solution of the COM is not independent of the eigenstate for relative wavefunction part.
In the weak confinement regime the overall energy of a state is $E_{nm}^{COM}=E_{r,n}+E_{COM,n,m}$ relative to the bandgap energy.

The weak confinement regime is more suitable for lower energy excitons in larger nanoplatelets. In this case, we may have to include the variation of the optical field over the nanoplatelet for the calculation of the nanoplatelets dipole moments.
For an in plane wave vector of the external optical field  $\mathbf{k}_{\parallel}$, the dipole strength is determined by the following integral \cite{Zimmermann::03}:
$
 O_{COM,nm}=\psi_{r,n}(\mathbf{r}=0) \int \mathrm{d}^2R e^{-\imath \mathbf{k}_{\parallel}\cdot \mathbf{R}}  \psi_{COM,n,m}(\mathbf{R}).$
To simplify the comparison to the strong confinement regime, we apply the limit $\mathbf{k}_{\parallel}\approx 0$ assuming small platelets compared to the wavelength of the radiation \cite{Zimmermann:PSSa:97}.

\section{Exciton states and absorption spectra}
In this section, we will calculate simple absorption spectra using the full solution and the approximate solutions for the weak and strong confinement case.
We use the formula $\alpha(\omega)=-\mathrm{Im}(\sum_{\{n\}}  |O_{\{n\}}|^2/(E_{\{n\}}-\omega+\imath \gamma))$ for the absorption spectrum using an artificially set broadening $\gamma$ and sum over all quantum numbers $\{n\}$ of the respective ansatz \cite{Haug::94}. 
This allows a quick estimation, how well optical spectra in the different limits are described by the respective ansatz.

The full four dimensional wave function $\Psi(\bm{\rho}_e,\bm{\rho}_h)$ is hard to visualize, accordingly for comparing the full wave function to the two approximate solutions, we project the four dimensional wave function $\Psi(\bm{\rho}_e,\bm{\rho}_h)$ to the coordinates used for the different factorizations of the approximations.
For a factorization in electron and hole part the projection is (exact within the strong confinement limit)
$ |\tilde{\psi}_e(\bm{\rho}_e)|^2=\int \mathrm{d}^2 \bm{\rho}_h |\Psi(\bm{\rho}_e,\bm{\rho}_h)|^2$ and
$ |\tilde{\psi}_h(\bm{\rho}_h)|^2=\int \mathrm{d}^2 \bm{\rho}_e |\Psi(\bm{\rho}_e,\bm{\rho}_h)|^2$.
Furthermore for a factorization in relative and center of mass coordinates (exact in weak confinement limit), the projection has the form:
  $|\tilde{\psi}_{COM}(\mathbf{R})|^2=\int \mathrm{d}^2 \mathbf{r} |\Psi(\mathbf{R},\mathbf{r})|^2$
 and $|\tilde{\psi}_r(\mathbf{r})|^2=\int \mathrm{d}^2 \mathbf{R} |\Psi(\mathbf{R},\mathbf{r})|^2$.
For optical transitions, the integral over the full wave function with electron and hole at the same position $\Psi(\tilde{\bm{\rho}},\tilde{\bm{\rho}})=\Psi(\mathbf{R}=\tilde{\bm{\rho}},\mathbf{r}=0)$ determines the oscillator strength for interband transitions, therefore plots over  $\Psi(\tilde{\bm{\rho}},\tilde{\bm{\rho}})$ help furthermore to understand the optical properties of the full solution.  

In supplemental material plots of projections of all calculated wave functions for different platelets are included together with absorption spectra calculated using the full solution and the approximate solutions for reference.

 \begin{figure}[htb]
  \includegraphics[width=8cm]{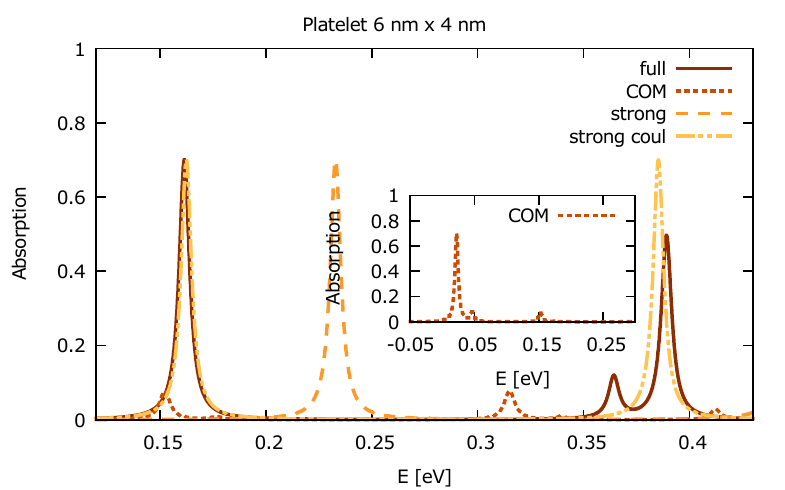}  
  \caption{Normalized calculated absorption spectrum for a $6~\mathrm{nm}\times4~\mathrm{nm}$ platelet (strong confinement).With the following nomenclature for the different absorption spectra: {\it full} for using the  full  exciton wavefunction, {\it COM } for the factorization into relative and center of mass coordinates, {\it strong} for factorization into electron and hole wavefunctions without Coulomb shift and {\it strong coul} for factorization into electron and hole wavefunctions including Coulomb shifts.}
  \label{abs_spec_strong}
 \end{figure} 

 \begin{figure}[htb]
\includegraphics[width=6.63cm]{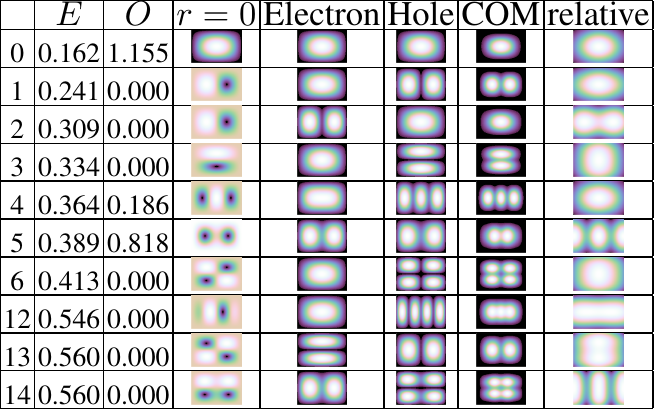}
\caption{Plots of selected exciton states and table of exciton energy $E$ in $eV$ and oscillator strength $O$ in arbitrary units (only comparable to the same platelet) for a $6~\mathrm{nm}\times4~\mathrm{nm}$ platelet, an example for the strong confinement case. Here, $r=0$ is a plot of $\Psi(\mathbf{R}=\tilde{\bm{\rho}},\mathbf{r}=0)$, which integrated over $\tilde{\mathbf{r}}$ determines the oscillator strength, {\it Electron} means the projection of the full wave function to the averaged electron wavefunction $|\tilde{\psi}_e(\bm{\rho}_e)|^2$, {\it Hole}, {\it COM}, and {\it relative} does the same for averaged hole  $|\tilde{\psi}_h(\bm{\rho}_h)|^2$, center of mass $|\tilde{\psi}_{COM}(\mathbf{R})|^2$ and relative wavefunction $|\tilde{\psi}_r(\mathbf{r})|^2$.}
\label{states_strong}
\end{figure}

\subsection{Strong confinement}
For discussing an example of the strong confinement case, we take a $6~\mathrm{nm}$ by $4~\mathrm{nm}$ platelet, which almost reaches strong confinement.
The calculated absorption spectrum in Fig. \ref{abs_spec_strong} uses exciton states obtained from the calculation of the full wave function and from calculations using 
 the different approximations. The absorption spectra provide a quick way to judge the quality of exciton energies and optical selection rules (resulting in the oscillator strengths) in the respective approximations.
The energy is always given relative to the band gap of the material, this allows to see immediately for the exciton ground state, if confinement (positive) or Coulomb binding energy (negative) dominates.

The spectrum of the full calculation shows mainly two peaks: an s-s exciton build up from an s-like electron and an s-like hole (cf. projected electron and hole wavefunctions for state $0$ in Fig. \ref{states_strong}) and an p-p exciton consisting of an p-like electron and p-like hole (see state $5$ in Fig. \ref{states_strong}).
As expected for strong confinement case with a confinement in the order of or below the exciton Bohr radius,
the factorized electron and hole wavefunction yields an overall good agreement only the oscillator strength of the higher energy p-p state is slightly overestimated in the approximate solution.
The  Coulomb coupling between electron and hole  in first order perturbation theory is sufficient to correct the s-s and p-p exciton energy  for the factorized electron and hole wavefunction ansatz. Without considering the Coulomb correction the energies of the strong confinement solution differ considerably. 
Despite the good agreement with the Coulomb corrected strong confinment, a small deviation between the full solution and factorized electron and hole wavefunction is visible: state $4$ has a small non negligible oscillator strength (cf. Fig. \ref{states_strong}) in the full solution as opposed to a vanishing oscillator strength, which would be expected for strong confinement regime. 
Looking at the averaged hole wavefunction  $|\tilde{\psi}_h(\bm{\rho}_h)|^2$ in Fig. \ref{states_strong} shows the reason: the factorized hole wavefunction in one dimension for an infinite deep quantum box is a sinus function having equally spaced zeros along the axis, but here the projected hole wave function has unequally spaced zeros caused by influence of the Coulomb interaction. Overall the Coulomb coupling changes  $\Psi(\mathbf{R}=\tilde{\bm{\rho}},\mathbf{r}=0)$ compared to the approximate solution, so that the overall oscillator strength does not  completely vanish for the state build up mostly from a s-like electron and a d-like hole. However the oscillator strength is weak enough compared to the other states, that it will be very hard to detect this state 
spectroscopically, if inhomogeneous broadening is present. 

As expected for the strong confinement case, the exciton states obtained from the COM ansatz do not describe at all the spectrum for the  $6~\mathrm{nm}$ by $4~\mathrm{nm}$ platelet.
The energy of lowest exciton state is off by several hundred meVs and the overall peak structure of the COM ansatz does not match the full solution. 
Now, we have  a closer look to the plots of the averaged full exciton wavefunctions in Fig. \ref{states_strong} of selected dark exciton states  between the two dominant bright excitons. (Full plots of all calculated exciton states can be found in supplemental material). First of all we see that the averaged relative wavefunctions $|\tilde{\psi}_r(\mathbf{r})|^2$ appear  blurry, this is a hint that a factorization into relative and center of mass parts is a  bad approximation in this limit.
Since the mass of the hole is higher than for the electron, the dark states consists mainly of states with holes with a higher orbital momentum. Only one exciton state energetically between the two bright states involves a p-like electron state.
Multiple dark intermediate states between the two optical bright state carry very different angular momentum, although angular momentum is not a good quantum number here. 
The presence of dark states and their angular momentum properties will be important for exciton-phonon relaxation studies, which will be subject to future studies.

  \begin{figure}[htb]
  \includegraphics[width=8cm]{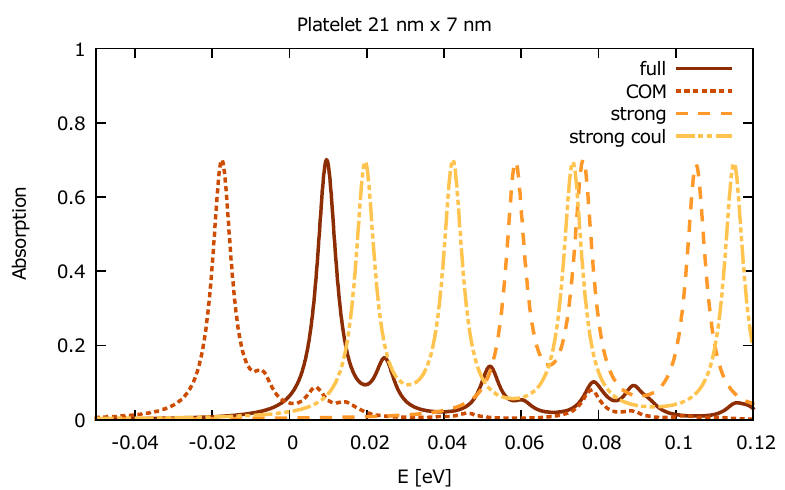}
  \caption{Normalized calculated absorption spectrum for 
  $21\mathrm{nm}\times7\mathrm{nm}$ platelet (weak and intermediate confinement). See caption of Fig. \ref{abs_spec_strong} for nomenclature.}
  \label{abs_spec_21_7}
 \end{figure} 
  
  \begin{figure}[htb]
  \includegraphics[width=8cm]{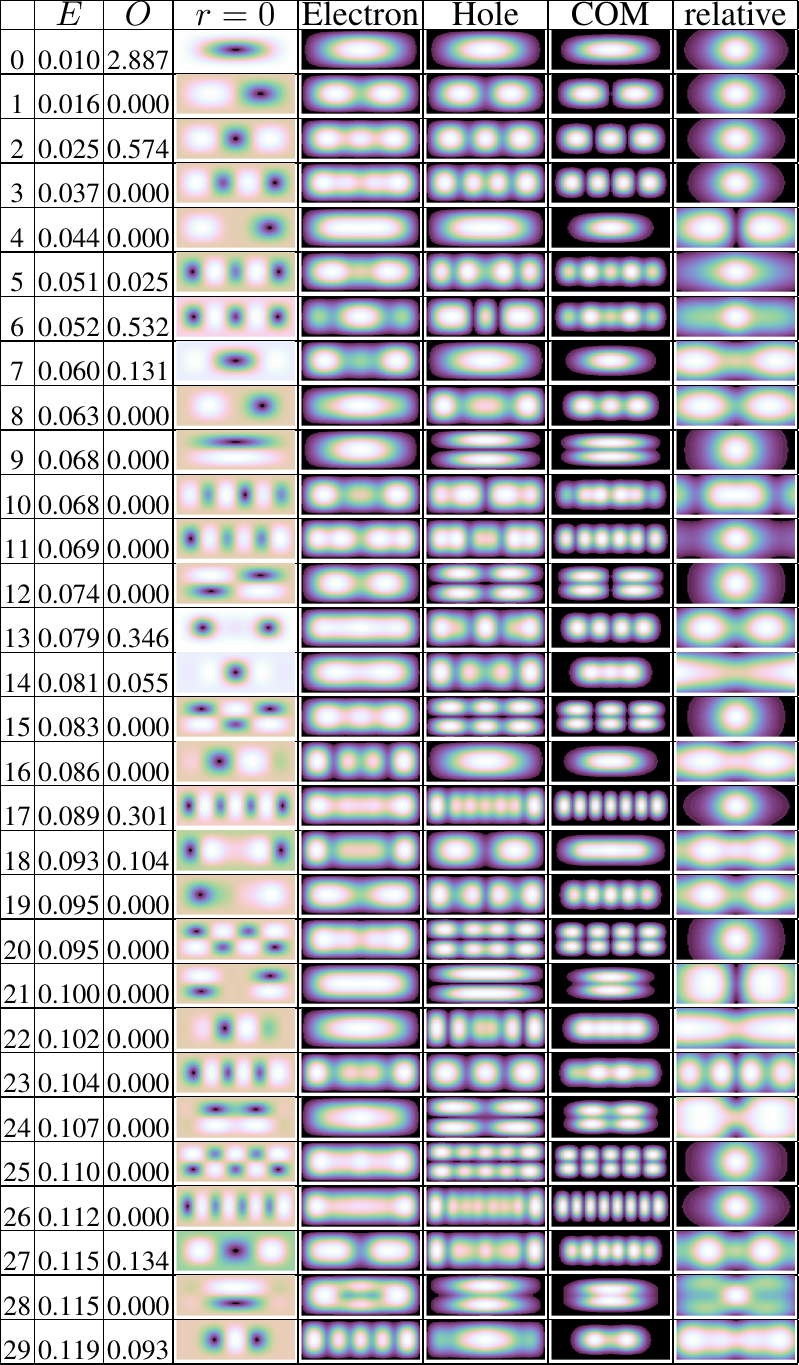}
  \caption{Plots of selected exciton states and table of exciton energy $E$ in $eV$ and oscillator strength $O$ in arbitrary units (only comparable to the same platelet)  for a $21\mathrm{nm}\times7\mathrm{nm}$ platelet, an example for the weak and intermediate confinement case. See Fig. \ref{states_strong} for nomenclature of the different columns.}
  \label{states_21_7}
 \end{figure}

\subsection{Intermediate and weak confinement}
In this section we discuss the intermediate and weak confinement regime.
 
We start with a $21~\mathrm{nm} \times 7~\mathrm{nm}$ platelet, which is larger than the previously discussed $6~\mathrm{nm}\times 4~\mathrm{nm}$ platelet. It is an example of  the intermediate regime between strong and weak confinement. 
The  $7~\mathrm{nm}$ length suggests that  strong confinement regime is close, 
while $21~\mathrm{nm}$ length suggests weak confinement.
The absorption spectra calculated using the full wave function and the approximations are plotted in Fig. \ref{abs_spec_21_7}.
The results from the strong confinement ansatz using factorized electron and hole wave functions show almost no similarity to the absorption spectrum calculated using the full calculation. Neither the overall distribution of bright states, oscillator strength nor the exciton energy  match the full solution. Only the energy of the lowest energy exciton of the exact result is not far from the Coulomb corrected strong confinement result.
The weak confinement approach using a factorization into center of mass (COM) $\psi_{COM}(\mathbf{R})$ and relative wavefunction $\psi_r(\mathbf{r})$
yields at least roughly qualitatively a similar arrangement of the peaks as the full solution.
However the COM exciton states are shifted towards lower energies compared to the full solution, because the com approach uses a free relative wavefunction resulting in a circular wavefunction for states with a s-type relative wavefunction $\psi_r(\mathbf{r})$. 
But in reality the relative wavefunction should be more elliptical (cf. the averaged relative wavefunction of the full solutions for  states 0-3 in Fig. \ref{states_21_7}). This deformation induced by the $7~\mathrm{nm}$ confinement length reduces the Coulomb binding energy. 
So for this platelet  a strong confinement approach in the smaller direction and a weak confinement approach in the larger direction may be a more appropriate ansatz.

We discuss the spectrum of the full solution in order to understand the remaining qualitative and quantitative differences between the full solution and the com solution: 
The full solution in Fig. \ref{abs_spec_21_7} shows four groups of peaks: one group around $0-0.02~\mathrm{eV}$, a second group around $0.04-0.06~\mathrm{eV}$, a third group around  $0.07-0.1~\mathrm{eV}$ and a fourth group beyond $0.1~\mathrm{eV}$.
The first group consists of states 0 and 2   (cf. Fig. \ref{states_21_7}), whose averaged relative wavefunction $|\tilde{\psi}_r(\mathbf{r})|^2$ looks like a deformed 1s exciton state, which explains the energy shift compared to the com solution with an undeformed 1s state in the relative part.
The averaged COM wave function $|\tilde{\psi}_{COM}(\mathbf{R})|^2$  shows one maximum for state 0 and three maxima for state 2. 
The second group of peaks consists of states 5, 6 and 7, their averaged relative wavefunction $|\tilde{\psi}_r(\mathbf{r})|^2$ looks like a superposition of 1s, 2s and 2p states, where 2s  contribution should be  bright and 2p contributions should be dark.
 The bigger mean radius of the 2s and 2p states compared to the 1s contribution cause the superposition of these states. 
 A superposition of 1s, 2s and 2p relative states can not be described with the simple COM approach. This causes here the deviation between the full solution and COM approach.
The third group (states 13, 14, 17, 18), fourth group (states 27 and 29) of bright states show similar features as the first two groups, the states include deformed relative 1s states (state 17) and higher energy excited p states (states 13, 14, 18). For higher energy excitons the averaged relative wavefunction form increasingly deviates from unconfined s-, p- and d-type functions due to the confinement area.
So the underestimated dipole moment and shifted exciton energy of higher energy states in COM solution compared to the full solution is caused by the disregard of confinement potential in the calculation of the relative  Coulomb dominated wavefunction.
However overall the qualitative agreement is acceptable for the COM solution for lower energy exciton states.
On the other hand our analysis clearly showed that the factorized approach is beyond its validity for the higher excited states in the platelet and the full solution is required. Since the radius of the relative wavefunction increases for higher excited states we will never find a platelet, that is completely inside the weak confinement regime. However depending on the platelet size more and more lower exciton states will enter the weak confinement regime.

The comparison between the full solution and the COM solution allows also to the review the validity of a rule of thumb: 
The exciton Bohr radius (connected to the classical radius of the 1s state) is often used   to discriminate the weak and strong confinement regimes.
We noticed that for the low energy 1s like states 0 and 2, the COM solution is a good first estimation even in the intermediate regime, since the averaged relative wave function is mainly smaller than the platelet box (at least in the 21 nm direction).
However for the higher energy excited states this is not true, since in the COM approach, the 2s and 2p states of the relative wavefunction supply another set of  bright (2s) and dark (2p) states for higher energy excitons. The averaged radius of the higher excited states  is much bigger than the 1s state and does not  at all fit within the platelet area, so the rule taking the 1s exciton Bohr radius (connected to the average radius by a factor) as measure does not apply for higher energy exciton states.

  \begin{figure}[htb]
    \includegraphics[width=8cm]{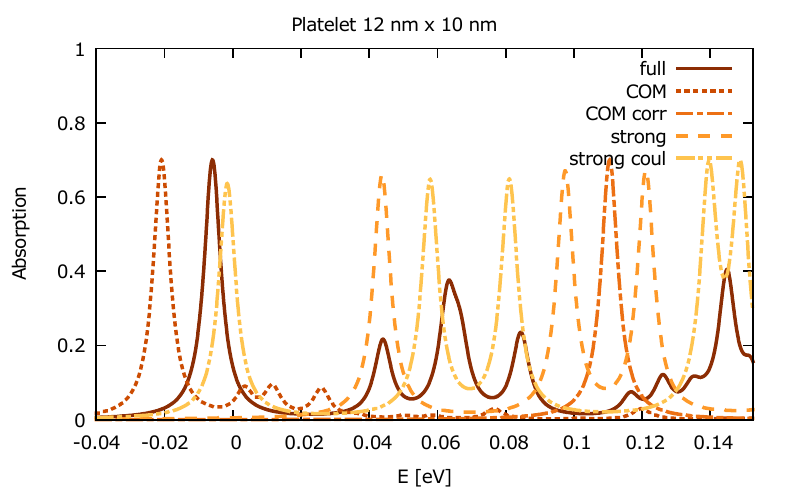}
  \caption{Normalized calculated absorption spectrum for 
  a $12\mathrm{nm}\times10\mathrm{nm}$ platelet (intermediate confinement). See caption of Fig. \ref{abs_spec_strong} for nomenclature.}
  \label{abs_spec_12_10}
 \end{figure} 
   
   \begin{figure}[htb]
  \includegraphics[width=6.7cm]{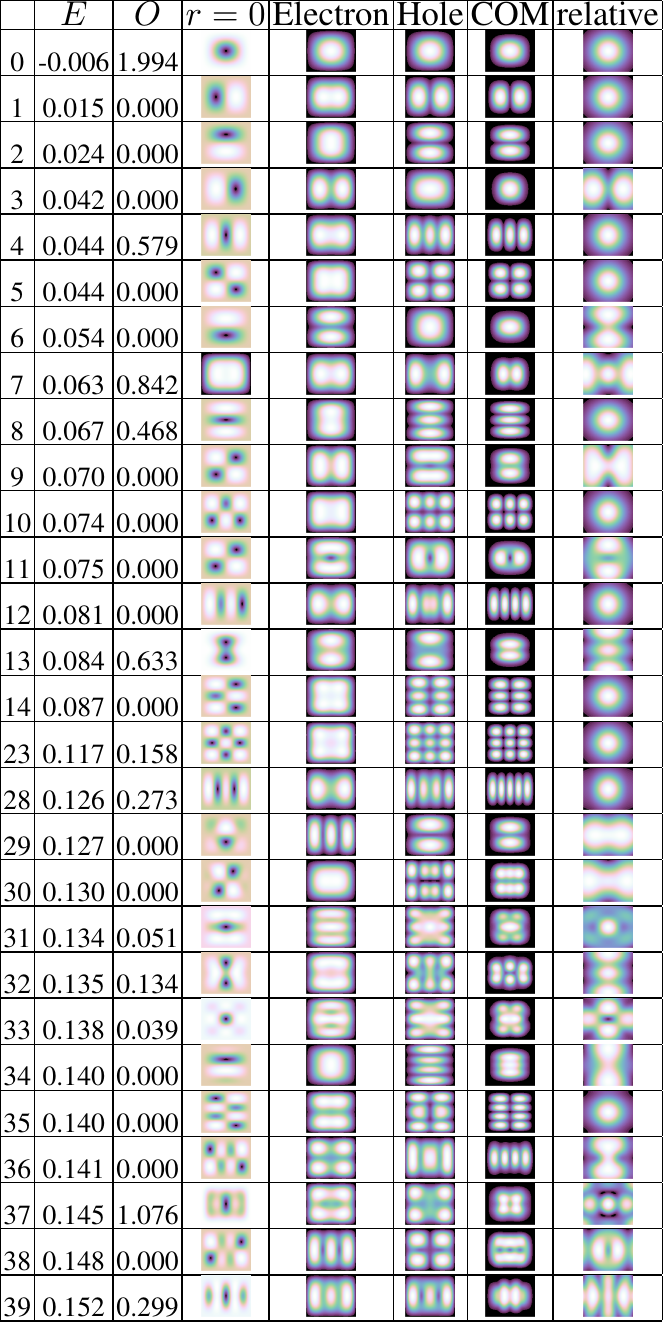}
  \caption{Plots of selected exciton states and table of exciton energy $E$ in $eV$ and oscillator strength $O$ in arbitrary units (only comparable to the same platelet) for an $12\mathrm{nm}\times10\mathrm{nm}$ platelet, an example for the weak to intermediate confinement case. See Fig. \ref{states_strong} for nomenclature of the different columns.}
  \label{states_12_10}
 \end{figure}
  \begin{figure}[htb]
    \includegraphics[width=8cm]{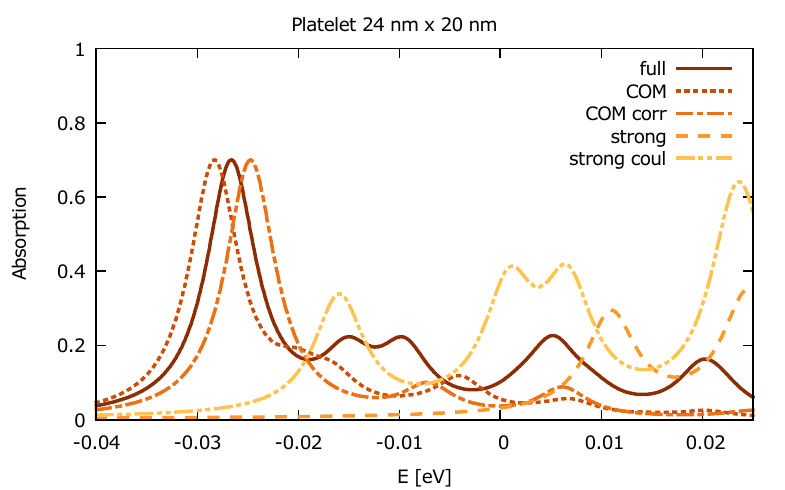}
  \caption{Normalized calculated absorption spectrum for $24\mathrm{nm}\times20\mathrm{nm}$(weak and intermediate confinement). See caption of Fig. \ref{abs_spec_strong} for nomenclature.}
  \label{abs_spec_24_20}
 \end{figure}
  \begin{figure}[htb]
  \includegraphics[width=6.7cm]{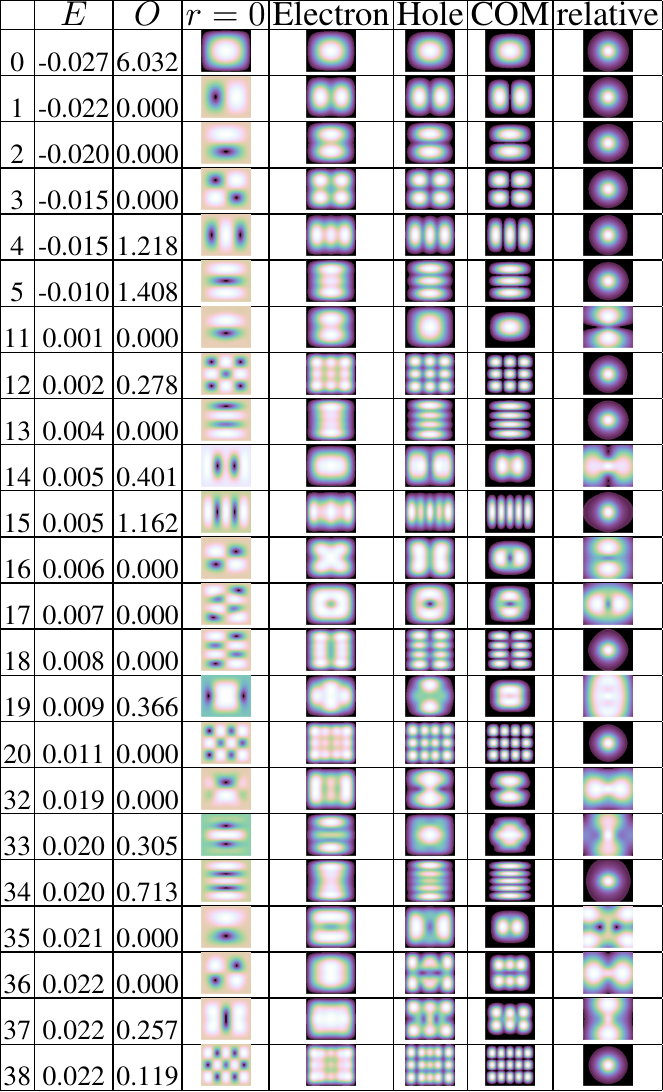}
  \caption{Plots of selected exciton states and table of exciton energy $E$ in $eV$ and oscillator strength $O$ in arbitrary units (only comparable to the same platelet)  for an $24\mathrm{nm}\times20\mathrm{nm}$ platelet, an example for the weak  confinement case. See Fig. \ref{states_strong} for nomenclature of the different columns.}
  \label{states_24_20}
 \end{figure}

For the $21~\mathrm{nm}\times7~\mathrm{nm}$ nanoplatelet, the $21~\mathrm{nm}$ length suggested weak confinement, while the $7~\mathrm{nm}$ length suggested strong confinement. We will now turn to nanoplatelets, which are more quadratic, so that is clearer, which limit is expected.
A $12~\mathrm{nm}\times10~\mathrm{nm}$ nanoplatelet should be in the intermediate regime but still close to strong confinement limit.
The energy of the lowest energy exciton state of the full solutions in the absorption spectrum in Fig. \ref{abs_spec_12_10} 
is well reproduced by the strong confinement case with Coulomb correction. But also the lowest energy state from the COM ansatz is not far off.
The lowest energy exciton state averaged electron $|\tilde{\psi}_e(\bm{\rho}_e)|^2$ and hole wavefunction $|\tilde{\psi}_h(\bm{\rho}_h)|^2$ (see Fig. \ref{states_12_10}, state 0) show, that it is an s-s state, i.e. it has s-type electron and hole wave function. The s-s type symmetry clearly matches the strong confinement solution for the lowest energy exciton.
For the strong confinement solution  of bright states, the electron wave function have to match the symmetry of the hole wavefunction (e.g. s-s, p-p, d-d states should be bright). 
Beside the s-s state 0, we find p-p states 7 at $0.063~ \mathrm{eV}$ and 13 at $0.084 ~\mathrm{eV}$ and higher angular momentum states 31-33 and 37, 39 around $0.11-0.15~\mathrm{eV}$ (cf. Fig. \ref{states_12_10}). 
The Coulomb corrected strong confinement solution reproduces the contribution of the bright states well (cf. Fig. \ref{abs_spec_12_10}) only the oscillator strength differs and the energies are slightly shifted.
However, if we inspect averaged electron $|\tilde{\psi}_e(\bm{\rho}_e)|^2$ and hole wavefunctions $|\tilde{\psi}_h(\bm{\rho}_h)|^2$ in Fig. \ref{states_12_10}, we can see   small deviations from a perfect p-shape for e.g. state 7 and 13.  This deviations are also visible in the plot of $\Psi(\mathbf{R}=\tilde{\bm{\rho}},\mathbf{r}=0)$, which  is connected to the overlap of electron and hole wavefunction  and determines the oscillator strength. Of course these deviations are caused by Coulomb interaction and the onset of the intermediate regime. Similar deviations from expected shapes are also visible for higher exciton states.
Besides the bright excitons matching states expected  from the strong confinement approach,
we see bright exciton state like state 4, 8, which should not be bright, if we were strictly in the strong confinement regime.
Here the averaged electron $|\tilde{\psi}_e(\bm{\rho}_e)|^2$ and hole wavefunction $|\tilde{\psi}_h(\bm{\rho}_h)|^2$ for states 4 and 8 show a s-type electron and a d-type hole and should not yield oscillator strength.
This is the same type of additional bright state, which was already present at the $6~\mathrm{nm}\times4~\mathrm{nm}$ nanoplatelet, but now these states have similar oscillator strength like the close p-p type states and cannot be ignored.
These additional bright states are  an additional sign of entering the intermediate regime.

An almost quadratic  $24~\mathrm{nm}\times20~\mathrm{nm}$ nanoplatelet should enter at least for the lower energy exciton states the weak confinement regime.
The two COM calculations (with and without area correction) have the most agreement
with the full calculation in Fig. \ref{abs_spec_24_20}. For the lowest energy exciton the calculation with area correction underestimates slightly the exciton binding energy, whereas the calculation without area correction overestimates slightly the exciton binding energy.
This means with a platelet of this size, we are leaving the range, where the COM approach without area correction is appropriate, but the platelet is still too small for a perfect approximation using an  area corrected COM calculation.
For higher energy exciton states, the two COM calculations show some similarity regarding their peak distributions and exciton energies to the full solution, but the result is not completely convincing. The oscillator strength is underestimated and the exciton energies are shifted especially for higher energies. An effect which we have already seen for the $21~\mathrm{nm}\times7~\mathrm{nm}$ platelet.
The strong confinement ansatz shows a significantly higher energy for the lowest exciton state than the full solution.
Furthermore the oscillator strength distribution differs qualitatively from the full solution.
However the overall peak structure shows some similarities, but we will see by inspecting the exciton states, that in the full solution none of them shows the properties of a exciton state separable in electron and hole wavefunction as it is expected for strong confinement.
Actually, this is a quite dangerous situation for yielding a proper interpretation, when using approximate techniques. 
Since someone using the strong confinement approach might get a similar spectrum as the full solution by slightly adjusting some material parameters, while the nature of the states differs completely.

Now, we discuss the bright exciton states of the $24~\mathrm{nm}\times20~\mathrm{nm}$ nanoplatelet in detail:
We have several bright exciton state with an averaged 1s relative wavefunction $|\tilde{\psi}_r(\mathbf{r})|^2$: (states 0, 4, 5, 12, 38, cf. Fig. \ref{states_24_20}). The averaged electron and hole wavefunctions $|\tilde{\psi}_e(\bm{\rho}_e)|^2$ and $|\tilde{\psi}_h(\bm{\rho}_h)|^2$ of state 4 looks roughly like a d-d state.
On the other hand, if this interpretation holds, the same arguments lead to the conclusion, that state 1 is a bright p-p state, but state 1 is dark. Furthermore $\Psi(\mathbf{R}=\tilde{\bm{\rho}},\mathbf{r}=0)$ of state 4 shows positive and negative peaks, this would not be the case with a bright exciton wavefunction separable in electron and hole part, which both have the properties of a d like state. These points clearly show that the wavefunction cannot described within the strong confinement limit. 

We have again higher energy excitons with an averaged relative wavefunction looking like a superposition of 1s, 2s and 2p states and higher angular momentum states  (states include 14, 15, 19, 33, 34, 37).
Again especially the size of 2p relative state $|\tilde{\psi}_r(\mathbf{r})|^2$ is larger than the platelet size, so that here the deviation from the COM result is caused by the confinement, which is not included in the calculation of the relative wavefunction in the COM approach. So many of the higher excited states are in an intermediate regime and cannot be described by the simple approaches from the strong or weak confinement approach. In general weak confinement is only achieved for the lower excited states.

\section{Conclusion}
In this paper we have recapitulated common toy models for nanostructures in the context of nanoplatelets.
Namely approximations using the typical approaches for strong and weak confinement were compared with results from the full four dimensional Schrödinger equation. We laid special emphasizes  on the higher excited states.
The analysis showed that nanoplatelets  can be in weak or strong confinement regime depending on their size. But there exist many examples, where the nanoplatelet is actually in an intermediate regime.
Weak confinement regime was always only achieved for the lower energy exciton states. Also the typical rule of thumb using the exciton Bohr radius for discriminating the weak and strong confinement, is only applicable for the lowest energy exciton states.
The qualitative simple model system used for the analysis showed, that approaches relying completely on either strong or weak confinement have to be used with care for quantitative and qualitative analysis, if the size of the platelets are varied. This especially true for higher excited states.
At least approaches, that  correct for the effects of the intermediate regime have to be used to compensate for the effects shown in this paper.

\begin{acknowledgments}
Support from the Deutsche Forschungsgemeinschaft (DFG) through SFB 787 is gratefully acknowledged.
 Alexander Achtstein, Riccardo Scott, Andrei Schliwa and Ulrike Woggon are knowledged for bringing the nanoplatelets to the authors attention and for insightful discussions. Andreas Knorr is acknowledged for fruitful discussions.
\end{acknowledgments}
%


\end{document}